\documentclass{emulateapj}

\newcommand{\msun}{$M_\odot$}
\newcommand{\kmps}{km\,s$^{-1}$}

\shorttitle{White dwarf core composition}
\shortauthors{Vennes et al.}

\begin{document}

\title{The Core Composition of a White Dwarf in a Close Double Degenerate System
\footnote{B\lowercase{ased on observations collected at the} E\lowercase{uropean} 
O\lowercase{rganisation for} A\lowercase{stronomical} R\lowercase{esearch in the} S\lowercase{outhern} 
H\lowercase{emisphere,} C\lowercase{hile under programme}
ID 086.D-0562.}
}

\author{S. Vennes$^{1,}$\altaffilmark{2}, A. Kawka$^{1,2}$}
\affil{$^1$ Astronomick\'y \'ustav, Akademie v\v{e}d \v{C}esk\'e republiky, Fri\v{c}ova 298, CZ-251 65 Ond\v{r}ejov, Czech Republic}

\altaffiltext{2}{Visiting Astronomer, Kitt Peak National Observatory, National Optical Astronomy Observatory, 
which is operated by the Association of Universities for Research in Astronomy (AURA) under cooperative 
agreement with the National Science Foundation.}

\begin{abstract}
We report the identification of the double degenerate system NLTT~16249 that comprises a normal, hydrogen-rich (DA) white dwarf and a
peculiar, carbon-polluted white dwarf (DQ) showing photospheric traces of nitrogen. We disentangled the observed spectra
and constrained the properties of both stellar components.
In the evolutionary scenario commonly applied to the sequence of DQ white dwarfs, both carbon and nitrogen would be dredged
up from the core. The C/N abundance ratio ($\approx 50$) in the atmosphere of this unique DQ white dwarf suggests
the presence of unprocessed material ($^{14}$N) in the core or in the envelope. 
Helium burning in the DQ progenitor may have terminated early on the red-giant branch after a mass-ejection event 
leaving unprocessed material in the core although current mass estimates do not favor the presence of a low-mass helium core. Alternatively,
some nitrogen in the envelope may have survived an abridged helium-core burning phase prior to climbing the asymptotic giant-branch.
Based on available data, we estimate a relatively short orbital period ($P\la13$ hrs) and
on-going spectroscopic observations will help determine precise orbital parameters.

\end{abstract}


\section{Introduction}

The presence
of a large concentration of carbon, detected via C$_2$ molecular bands, in the otherwise helium-rich atmosphere of many old
($\gtrsim 10^9$ years) white dwarfs has been viewed for many years \citep[e.g.,][]{vau1979,koe1982} as evidence of the carbon-rich nature of
their cores: The material diffuses upward, away from the core, and is dredged-up to the surface by the deep helium convection zone developing in cool white
dwarfs \citep{fon1984,pel1986,mac1998}.
These objects, collectively known as DQ white dwarfs, form a homogeneous sequence of stars with surface
temperatures ranging from $\sim11,000$ to $\sim5,000$ K and with
temperature-correlated surface carbon abundances between $10^{-2}$ and $10^{-7}$ relative to
helium by number \citep{duf2005,koe2006}.

Evolutionary models \citep[e.g.,][]{sch1992} show that stars of low to intermediate masses, i.e., below 8 to 10 \msun,
end their nuclear burning history with a core composed primarily of carbon and oxygen or
heavier elements (Ne, Mg) in the upper mass range \citep{gar1997}.
In the past, spectroscopic searches for oxygen (ultraviolet CO bands) or nitrogen (violet CN bands) have
yielded abundance limits O/C$\approx0.1-1$ increasing with temperature \citep{koe1982} or O/C,N/C$\approx 10^{-2}-10^{-3}$ \citep{weg1984}. 
Although oxygen is a primary product of the triple-alpha (3-$\alpha$) nuclear burning process,
the actual amount of oxygen dredged-up along with carbon from the interior depends on its abundance profile in the core \citep{mac1998}.
On the other hand, nitrogen should be fully exhausted along with helium by the 3-$\alpha$ process
and is not expected in the atmosphere of white dwarfs with a fully-evolved core.
Recently, \citet{gae2010} reported the discovery of two oxygen-rich white dwarfs
alongside the carbon-rich DQ sequence. The measured O/C abundance ratios ($\log{\rm O/C}\approx 0.6-1.8$) imply
core compositions dominated by oxygen (80 to 99\% by mass). The stars were suspected by \citet{gae2010} to be the descendents of
the most massive stars to avoid core collapse \citep[9-10\msun;][]{gar1997}. 

Clearly, the surface composition reflects the white dwarf core composition which, in turn, is determined by the 
evolutionary path followed by the progenitor. In this context,
we present in Section 2 spectroscopic observations of the peculiar object NLTT~16249 that show a strong Balmer line series 
along with C$_2$ Swan and CN violet band absorptions. The observed radial velocity variations and a successful spectral
decomposition and model atmosphere analysis (Sections 3.1-3.3) show that the object is in fact a close double degenerate system comprising 
a normal DA white dwarf and a peculiar DQ of similar
optical luminosities. We examine possible evolutionary scenarios in Section 3.4, and we propose that
the photospheric carbon to nitrogen abundance ratio (C/N$\approx50$) indicates that the core
material was left partially unprocessed or that burning ceased before destroying all nitrogen in the envelope. 
Finally, we summarize in Section 4.

\section{Observations and classification}

We first observed NLTT~16249 on UT 2010 Mar 26 with the R-C spectrograph attached to the 4m
telescope at Kitt Peak National Observatory. We obtained three
consecutive exposures of 900~s each using the KPC-10A grating with the WG360 order blocking filter.
The spectra cover a wavelength range from 3660 to 6790 \AA\ with a dispersion of 2.8 \AA\ per pixel. This low dispersion
spectrum ($\lambda/\Delta\lambda \approx 1000$) revealed a spectroscopic combination that would be unusual in any star.
The hydrogen Balmer line series are blended with C$_2$ Swan bands \citep[the $d^3\Pi_g-a^3\Pi_u$ system; see][]{tan2007} 
and what appears to be 
the  $\Delta v = 0$ sequence of the CN violet band near 3880 \AA\ \citep[the $B^2\Sigma^+-X^2\Sigma^+$ system; see][]{ram2006}.

Next, we obtained a single echelle spectrum on UT 2010 Nov 7 (2400~s exposure) with
the X-shooter spectrograph \citep{ver2011} attached to the UT2 (Kueyen) at Paranal Observatory.
The spectrum covers a wavelength range from 3000 \AA\ to 2.5 $\mu$m on three separate arms
with a resolving power $\lambda/\Delta\lambda \approx 9100$ for the UVB arm, 8800 for the VIS, and 6200 for the NIR.
This intermediate dispersion spectrum 
confirmed our earlier findings and also revealed the presence of additional sequences from the $B-X$ CN system.

The star could be identified as a DQAZ, i.e., a helium-rich white dwarf with traces of carbon and hydrogen along
with additional elements such as nitrogen, although the strength of the hydrogen lines is incompatible with this interpretation.
More likely, NLTT~16249 is a DA+DQ(N) double degenerate system comprised of
a normal hydrogen-rich (NLTT~16249B) and a nitrogen-contaminated DQ white dwarf (NLTT~16249A). We propose a non-standard classification 
for the DQ(N) white
dwarf owing to the unforeseen presence of nitrogen \citep[for the current classification scheme see][]{sio1983}. 
The diagnostics employed to confirm the second possibility
are, first, a radial velocity study and, next, a spectral decomposition into two components.

\begin{figure}
\begin{center}
\includegraphics[width=0.49\textwidth]{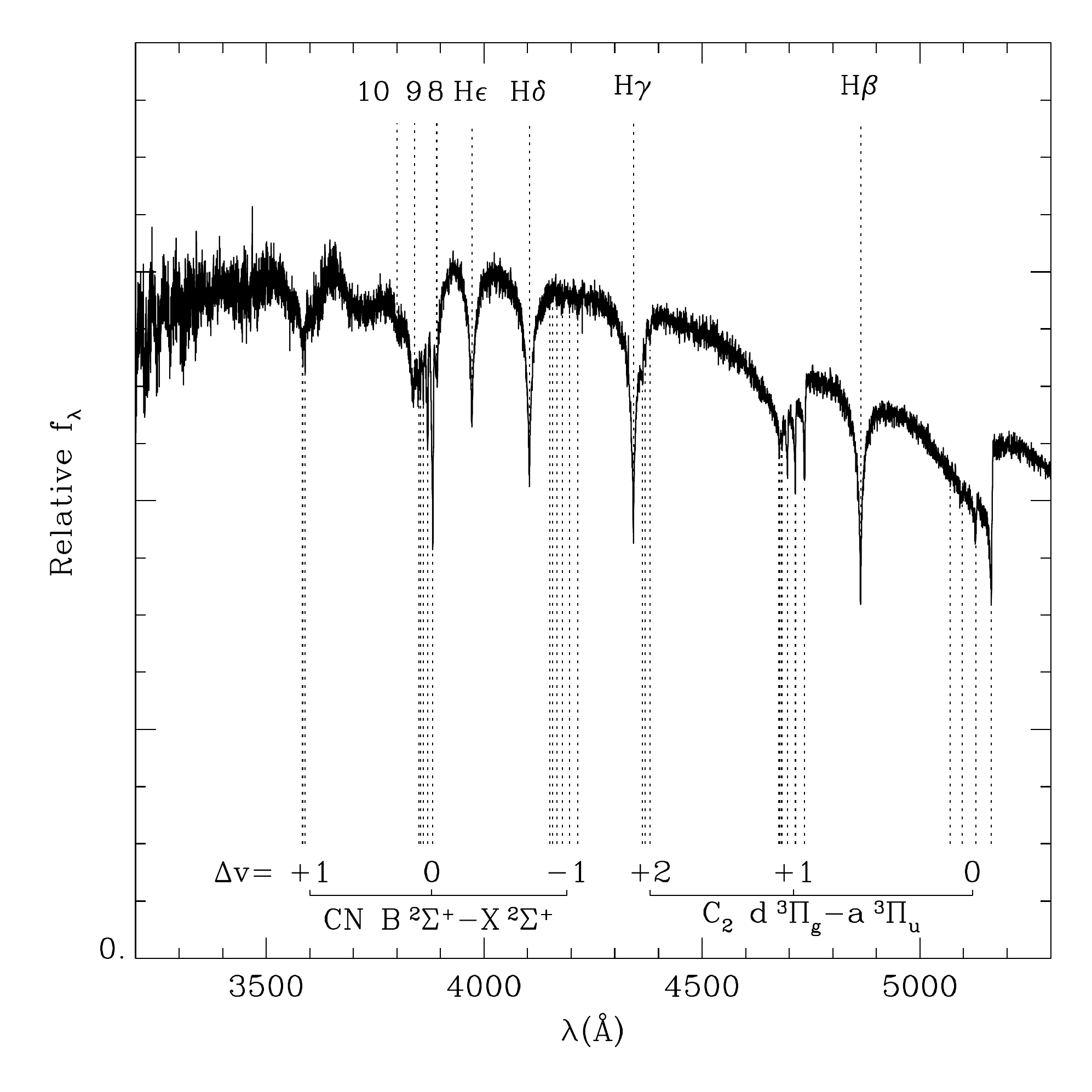}
\end{center}
\vspace{-0.5cm}
\caption{Spectrum of NLTT~16249 obtained with the UVB arm of X-shooter. The main spectral features
(Balmer lines and C$_2$ or CN molecular bands) are marked with dashed lines and labelled accordingly (see text).\label{fig_spec}}
\end{figure}

\begin{table}
\caption{Adopted C$_2$ and CN bandhead and $gf$ values \label{tbl-band}}
\centering
\begin{tabular}{rcclccl}
\hline\hline
           &            & \multicolumn{2}{c}{C$_2$} & & \multicolumn{2}{c}{CN} \\
	   \cline{3-4} \cline{6-7} \\
$\Delta v$ & $(v',v'')$ & $\lambda$\tablenotemark{a} (\AA) & $gf$ \tablenotemark{b} & & $\lambda$\tablenotemark{a} (\AA) & $gf$ \tablenotemark{c} \\
\hline
$-1$       &(0,1) & 5635.5 & 0.0198 & & 4216.0 & 0.00497 \\
           &(1,2) & 5585.5 & 0.0231 & & 4197.2 & 0.00826 \\
           &(2,3) & 5540.7 & 0.0192 & & 4181.0 & 0.0103  \\
           &      & &        & & &         \\
0          &(0,0) & 5165.2 & 0.0783 & & 3883.4 & 0.0684  \\
           &(1,1) & 5129.3 & 0.0300 & & 3871.4 & 0.0560  \\
           &(2,2) & 5097.7 & 0.0123 & & 3861.9 & 0.0466  \\
           &      & &        & & &         \\
$+1$       &(1,0) & 4737.1 & 0.0300 & & 3590.4 & 0.00767 \\
           &(2,1) & 4715.2 & 0.0435 & & 3585.9 & 0.0132  \\
           &(3,2) & 4697.6 & 0.0435 & & 3583.9 & 0.0170  \\
           &      & &        & & &         \\
$+2$       &(2,0) & 4382.5 & 0.0045 & & ... & ... \\
           &(3,1) & 4371.4 & 0.0102 & & ... & ... \\
           &(4,2) & 4365.2 & 0.0130 & & ... & ... \\
\hline
\end{tabular}
\tablenotetext{a}{From \citet{wal1962}.}
\tablenotetext{b}{From \citet{zei1987}.}
\tablenotetext{c}{From \citet{bau1988}.}
\end{table}

Figure~\ref{fig_spec} shows the X-shooter spectrum of NLTT~16249 (UVB arm): The main molecular bands are marked
at selected wavelengths \citep{wal1962} following standard spectroscopic notations along with the Balmer lines. 

The relative velocity change between the two components was readily measurable by using spectra obtained at the 
two available epoches, although individual velocity changes
may suffer from large uncertainties on the zero-point of the low-dispersion KPNO wavelength scale.  
The relative velocity change between the two components is free of systematic errors and allowed us to estimate
$K_{\rm DQ}+K_{\rm DA} \ga 320\pm30$ \kmps. On the other hand, the velocity shift of the DQ component was
$v_{\rm 1,DQ} - v_{\rm 2,DQ} + C = 120\pm30$\,\kmps,
where the subscripts ``1'' and ``2'' refer to the KPNO and VLT spectra, respectively, and $C$ is the zero-point error. Similarly, for the DA component
$v_{\rm 1,DA} - v_{\rm 2,DA} + C = -200\pm30$\,\kmps, i.e., in opposite direction to the DQ component. 
The amplitude ratio would imply that the DA white dwarf is more massive than its companion with 
$M_{\rm DQ}/M_{\rm DA}=1.7^{+0.9}_{-0.6}$, but because of potential systematic errors we will simply assume
that $M_{\rm DQ} \approx M_{\rm DA}$.

The binary has a significant proper motion of
$\mu_\alpha\cos{\delta}=2\pm6,\ \mu_\delta=-140\pm6$ mas yr$^{-1}$ \citep{sal2003} in a relatively crowded field. 
We collected Johnson photometry ($V=15.77$, $B-V=0.24$, $U-B=-0.66$ with an assumed error of 0.02 on the colors) from \citet{egg1968} 
and 2MASS photometry ($J=14.87\pm0.04$) from \citet{skr2006}. A close examination of the VLT acquisition image (epoch 2010)
shows a fainter crowded star to the NW. Taking into account the proper motion of NLTT~16249, the stars would have nearly overlapped circa 1990. 
The 2MASS-$J$ frame (epoch 1997) clearly shows a SE-NW elongation. Also, the DSS1 (epoch 1955) red and blue plates show that the crowded star is 
markedly red so that it probably contaminated the 2MASS-$J$ photometric measurement. With the crowded star at a larger separation, the UBV photometry obtained by
\citet{egg1968} circa 1965 was probably not significantly contaminated. In order to extend available broadband indices further in the red,
we folded the X-shooter UVB and VIS spectra with VRI bandpasses \citep{bes1990} and measured $V-R=0.15$ and $V-I=0.37$.

\begin{figure*}
\begin{center}
\includegraphics[viewport=0 0 580 580, clip, width=0.46\textwidth]{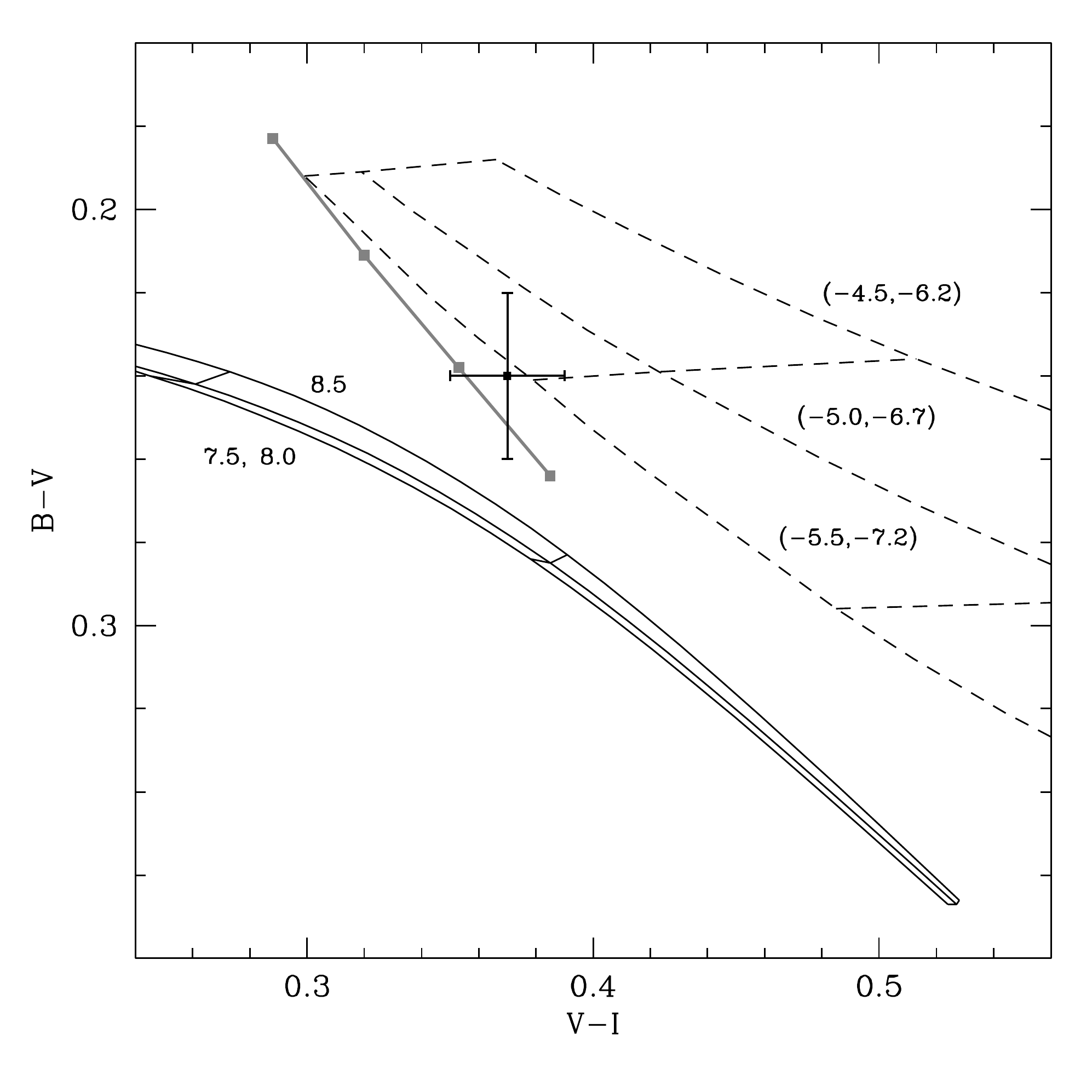}
\includegraphics[viewport=0 0 580 580, clip, width=0.46\textwidth]{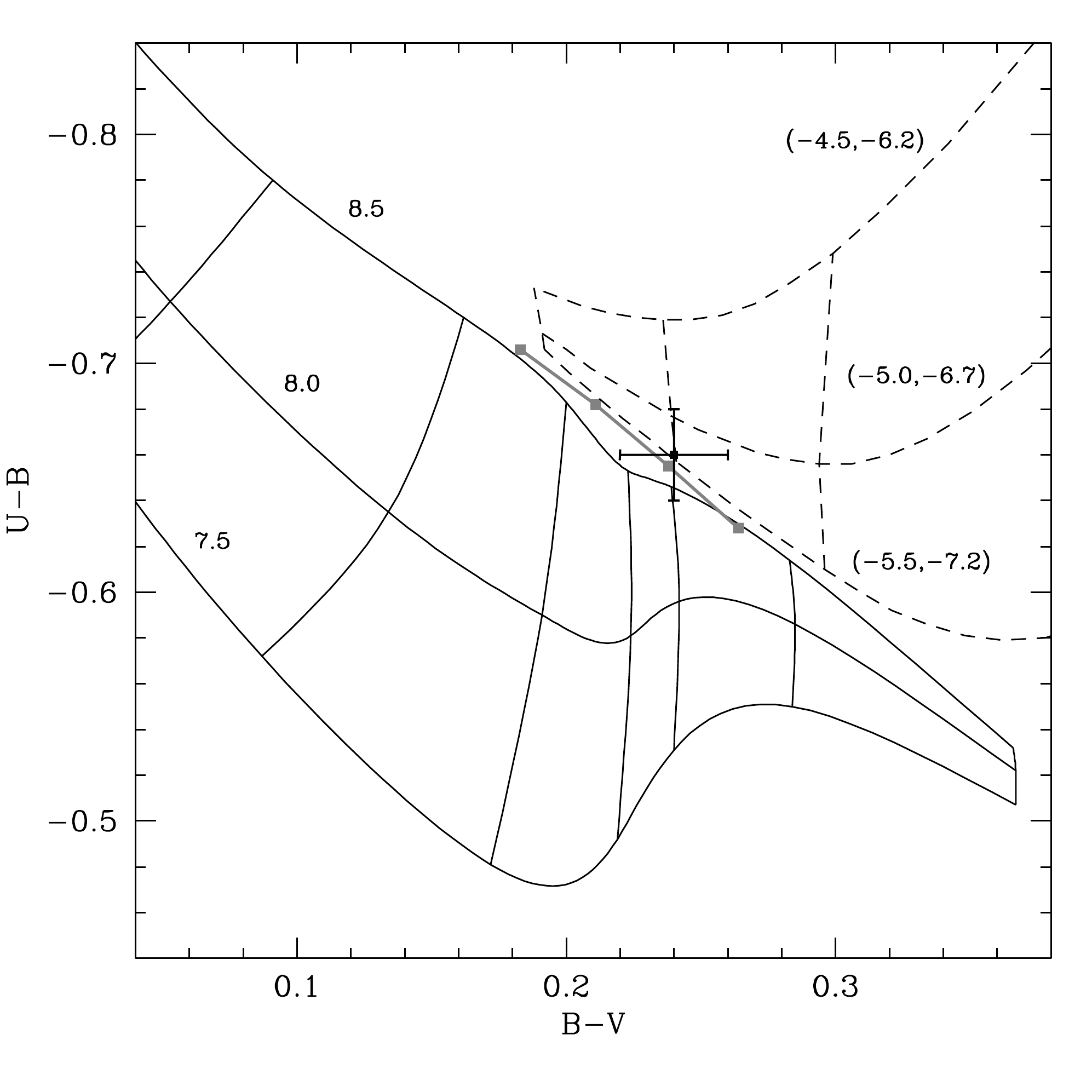}
\end{center}
\vspace{-0.5cm}
\caption{Johnson-Cousin color indices $B-V$ versus $V-I$ (left panel) and $U-B$ versus $B-V$ (right panel)
for pure-hydrogen (DA) model grids (full lines) and DQ model grids (dashed lines). The DA model grids are 
shown with varying surface gravity (labelled at $\log{g}=7.5, 8.0$, and $8.5$) and temperature (shown from right to left at 7, 8, 9, 10, 12, 14, 16$\times10^3$\,K).
The DQ model grids are shown with varying
carbon and nitrogen abundance (labelled with the pair $\log{\rm C/He},\log{\rm N/He}$) and temperature (shown from right to left at 7, 7.5, 8$\times10^3$\,K)
at $\log{g}=8$. On both panels the observed colors are shown with error bars, and the best-fit model colors are shown, from upper-left to lower-right, 
with grey squares and lines at $(\log{\rm C/He},\log{\rm N/He}) = (-4,-5.7),\ (-4.5,-6.2),\ (-5.0,-6.7),\ (-5.5,-7.2)$.
\label{fig_color}}
\end{figure*}

\section{Analysis}
 
The components were analyzed conjointly with the total flux measured at earth given by
\begin{equation}
f_\nu = \frac{4\pi}{D^2} \Big{(} R^2_{\rm DQ}\,H_{\rm \nu,DQ} + R^2_{\rm DA}\,H_{\rm \nu,DA} \Big{)}
\end{equation}
where $D$ is the distance to the binary, $R$ their respective radii, and $H_{\nu}$ their respective model Eddington fluxes.
The individual radii, hence masses, are constrained by the surface gravity of each model grid point using mass-radius relations
from \citet{ben1999}.
The best-fit parameters ($T_{\rm eff,DQ}$, $\log{g_{\rm\, DQ}}$, $T_{\rm eff,DA}$, $\log{g_{\rm\, DA}}$, and DQ abundances C/He, N/He) 
are obtained using $\chi^2$ minimization techniques.  We also computed Johnson-Cousin UBVRI magnitudes and color indices of the composite
spectra to constrain the best-fit parameters.
Because of the absence of CO spectroscopic signatures we set the oxygen abundance to zero in the DQ models. 
Moreover, the absence of Ca H\&K and CH G-band also suggests very
low metallicity in either star and the absence of hydrogen in the DQ white dwarf. With the revealing exception of nitrogen 
the DQ component appears normal.

\subsection{Model atmospheres}

The DA model atmospheres are described in \citet{kaw2006,kaw2011}.
The new DQ model atmospheres are in dual convective/radiative equilibrium as well as in local thermodynamic equilibrium. 
We included all relevant species (He, He$^+$, C, C$^+$, N, N$^+$, O, O$^+$) and molecules (C$_2$, CN, CO)
in the charge conservation and abundance equations: This non-linear system of equations was solved using the secant method
before each model iteration. We employed the molecular partition functions of \citet{sau1984}.
We also estimated the contributions of C$_2^+$ and concluded that virtually all electrons are contributed 
by the ionization of helium and carbon. 

The models add all relevant opacities, including ultraviolet \ion{C}{1} lines, 
and the C$_2$ Swan bands ($\Delta v=-1,0,+1,+2$) and the violet CN bands ($\Delta v=-1,0,+1$) using the ``just-overlapping line approximation''
\citep{gol1967,zei1982}. 
We also added the He$^-$ free-free and the C$^-$ bound-free and free-free opacities as well as helium Rayleigh scattering.
We updated the \cite{zei1987} CN molecular opacities 
with oscillator strengths from \citet{bau1988} and molecular constants tabulated in \citet{red2003} in good agreement
with \citet{ram2006}. Table~\ref{tbl-band} lists adopted bandhead wavelengths and $gf$ values for the first three vibrational transitions of the C$_2$ and CN systems. 

Figure~\ref{fig_color} shows relevant photometric properties of the model grids. In particular, optical DQ colors are sensitive to carbon (and nitrogen)
abundance because of varying depths of Swan and CN violet bands. This effect influences the composite colors and the selection of best-fit composite models.
We found that the DQ spectral energy distribution is only mildly affected by surface gravity variations \citep[see][]{duf2005}.
The DQ synthetic colors are markedly bluer than those of a DA model at the same temperature. 

\subsection{Properties of the components}

\begin{figure*}
\begin{center}
\includegraphics[viewport=0 0 580 580, clip, width=0.46\textwidth]{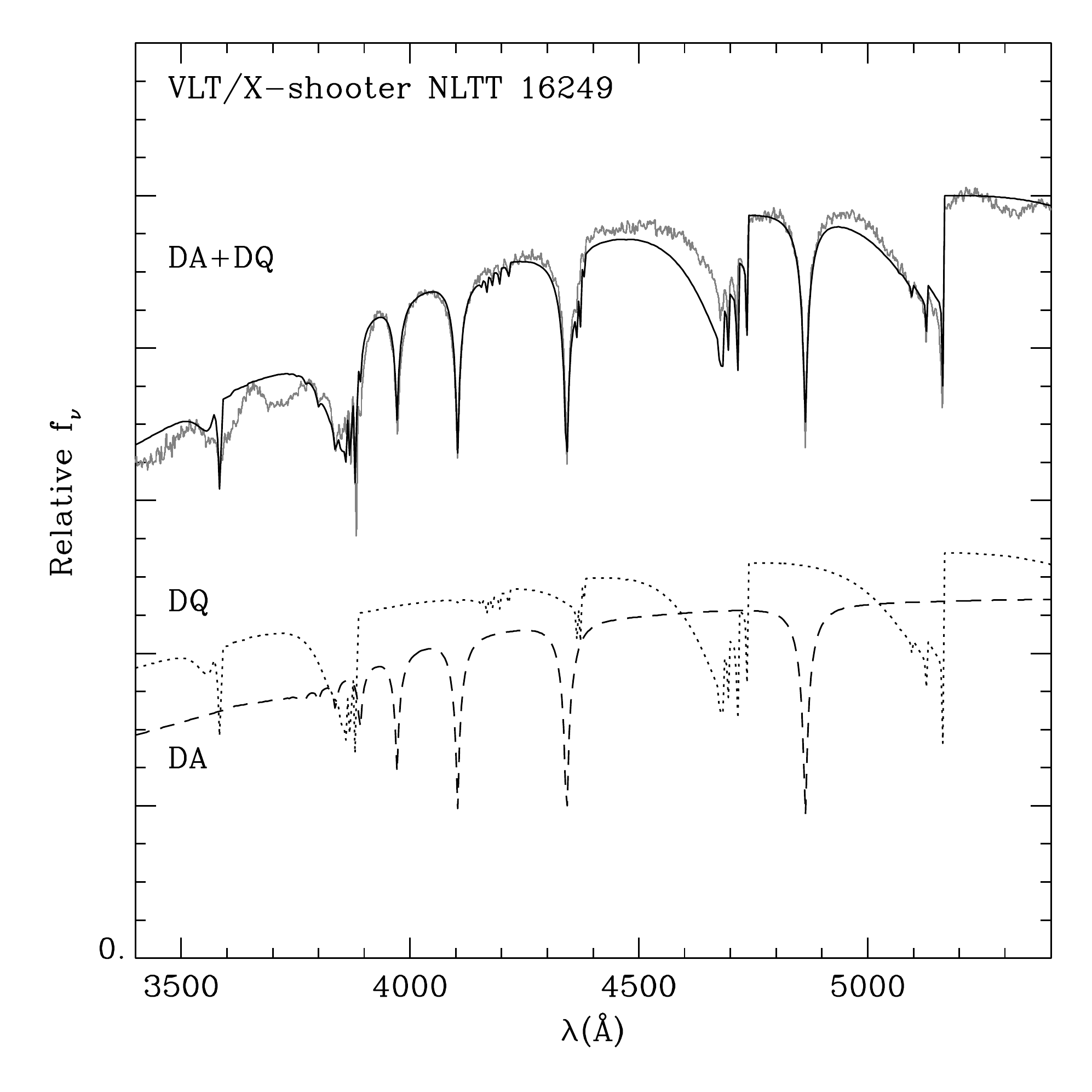}
\includegraphics[viewport=0 0 580 580, clip, width=0.46\textwidth]{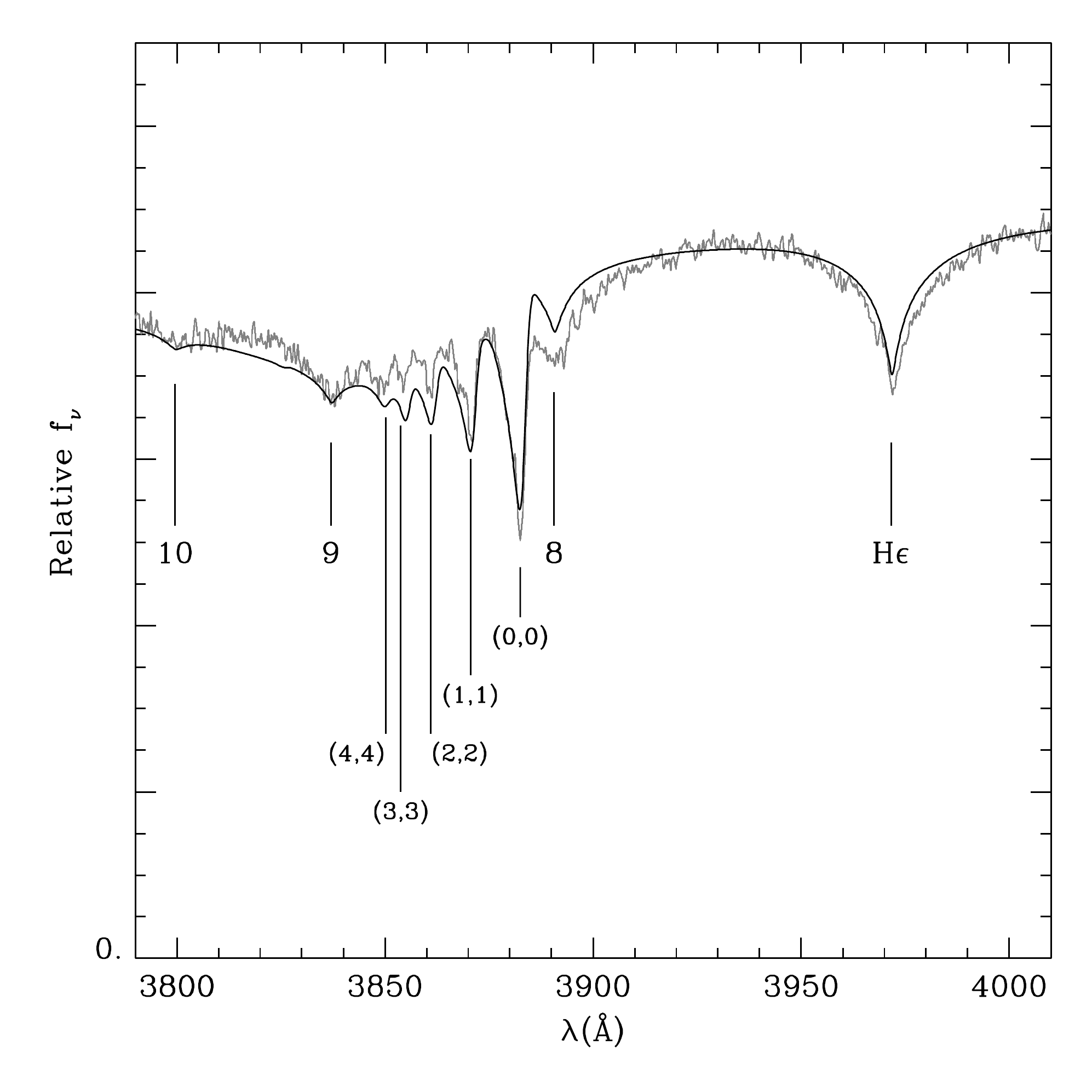}
\end{center}
\vspace{-0.5cm}
\caption{(Left panel) 
Model decomposition of the observed spectrum (degraded to a resolution of 2.4\AA\ for clarity). The model flux contributions of the DA and
DQ components are shown with dashed and dotted lines, respectively.
The total DA$+$DQ model flux (full line) is compared to the observed flux (grey line). 
(Right panel) 
Composite model spectrum fit to the X-shooter spectrum (full resolution) near the Balmer series limit
and the CN violet band ($\Delta v = 0$).
\label{fig_comp}}
\end{figure*}

First, we proceed with the spectral decomposition. We mapped the minimum $\chi^2$ values in the ($\log{g_{\rm\, DQ}}, \log{\rm C/He}$) plane corresponding 
to the best-fit DA parameters ($T_{\rm eff,DA}, \log{g_{\rm\, DA}}$) and DQ temperature ($T_{\rm eff,DQ}$). We varied $\log{\rm C/He}$ from $-5.5$ to $-4.0$
and $\log{g_{\rm\, DQ}}$ from 7.5 to 8.75. In all spectral decompositions the
nitrogen abundance scaled with the carbon abundance with a ratio $\log{\rm C/N}\approx 1.7$ that we subsequently held fixed at this value. 
The calculations show that a family of solutions
exists along an axis in the ($T_{\rm eff,DQ}, \log{\rm C/He}$) plane where higher DQ temperatures are compensated with higher carbon abundances.
This degeneracy of the solutions may be lifted by comparing the predicted colors of the model composites with observed colors. On the other hand,
the minimum $\chi^2$ values in the ($\log{g_{\rm\, DQ}}, \log{\rm C/He}$) plane are steadily found for $\log{g_{\rm\, DA}}\approx 8.2-8.5$. Interestingly,
the best-fit surface gravities (hence radii) always indicated similar binary component luminosities, so that the DQ gravity correlates
with the DA gravity and was found in the range $\log{g_{\rm\, DQ}}\approx 8.0-8.5$.

The predicted color indices were compared to observed indices
to constrain possible solutions.
Figure~\ref{fig_color} shows composite colors corresponding to the best-fit models with $\log{\rm C/He}$ varying from $-5.5$ to $-4.0$. 
As noted above, higher carbon abundances are compensated with high DQ temperatures. Both sets of indices favor a lower carbon abundance
\begin{displaymath}
\log{\rm C/He}=-5.0\pm0.3,\ \log{\rm N/He}=-6.7\pm0.3,
\end{displaymath}
and we conclude that the DA model parameters are
\begin{displaymath}
T_{\rm eff,DA} = 8330\pm200\,K,\ \log{g_{\rm\, DA}}=8.37\pm0.15,
\end{displaymath}
and, correlating with the DA parameters, the DQ parameters are
\begin{displaymath}
T_{\rm eff,DQ} = 7770\pm200\,K,\ \log{g_{\rm\, DQ}}=8.27\pm0.22.
\end{displaymath}
The total system mass varies between 1.33, in the lower gravity range, to 1.83\msun, in the higher gravity range. 
The distance calculated using the $V-M_V$ modulus varies from $\sim$34 pc for the high system mass to $\sim$46 pc for the low system mass. A parallax measurement would help confirm and narrow down the range of spectroscopic solutions.

Assuming a system mass in the range $M_{\rm DA}+M_{\rm DQ}=1.3-1.8$\msun\ and a minimum velocity amplitude $K_{\rm DA}+K_{\rm DQ}\ga 320$ \kmps, the orbital period is constrained to 
\begin{displaymath}
P = 2\pi G\,\frac{M_{\rm DA}+M_{\rm DQ}}{(K_{\rm DA}+K_{\rm DQ})^3}\,\sin^3{i} \la 9.2-12.8\,{\rm hrs}.
\end{displaymath}
Owing to the large velocity amplitude and relatively short period, the orbital parameters ($P,\,e,\,K_{\rm DA},\,K_{\rm DQ}$) should be easy to determine.

Figure~\ref{fig_comp} shows the best-fit composite model spectrum to the X-shooter spectrum using parameters listed above.
The Swan $\Delta v= +1$ bands appear too strong relative to $\Delta v=0$ bands, particularly
in the extended wing. 
Fortunately, the abundance offset between these two bands does not exceed $\approx 0.2$ dex.

\subsection{The dominant carbon isotope}

The isotopic shift of the $\Delta v=+1$ band of $^{13}$C$^{12}$C relative to that of C$_2$ is 7.5 \AA\ \citep{weg1984a}. By attributing the existing
band to C$_2$ we conclude that the absence of red-shifted absorptions implies C$_2$/$^{13}$C$^{12}$C$\,\ga\,$10,
i.e., $^{12}$C/$^{13}$C$\,\ga\,$10. 

\subsection{Evolutionary scenarios}

Hydrogen-deficient white dwarfs \citep[DB and DQ white dwarfs, see][]{sio1983} are possibly the product
of a late thermal pulse. This process leads to the formation of a hydrogen-deficient surface that is
also enriched in carbon and oxygen \citep[][and references therein]{her1999}. 
After reaching the white dwarf cooling sequence these objects develop a nearly pure-helium surface and 
are recognized as DB white dwarfs. While cooling further they develop a deep helium-dominated convection zone
dredging-up carbon from the core \citep{fon1984,pel1986,mac1998}.

The material mixed in the helium-dominated convection zone of the DQ white dwarf NLTT~16249A is enriched with
carbon and nitrogen. The diffusion time-scales at the bottom of the helium convection zone of a $\sim 8000$ K 
white dwarf are relatively short \citep[$\approx 10^3$ yr;][]{koe2009}. Therefore, a steady-state abundance of nitrogen and
carbon is achieved only if the carbon/nitrogen-rich material is supplied to the envelope from the core. This scenario,
commonly applied to ordinary carbon-rich DQ white dwarfs, should also apply to the peculiar DQ NLTT~16249A, but the source
of nitrogen remains elusive.

A copious amount of nitrogen is produced in intermediate-mass stars between
pulses on the AGB when the convective envelope reaches
the hydrogen-burning shell \citep{lat1996}. This production is deemed
responsible for nitrogen enrichment in planetary nebulae \citep{kar2009},
but what fraction of it would remain within the helium envelope and contribute
to the white dwarf core composition is an open question.

Nitrogen leftover material is present in the early-stages of helium burning on the horizontal-branch (HB), but it is
quickly destroyed along with helium leading to the formation of a carbon/oxygen-rich core. For example, a carbon-to-nitrogen
ratio of $\approx 50$ is briefly achieved in the core of 2 \msun\ HB star after $\sim 110$ Myr, but it rapidly declines afterward \citep{sch1992}. 
At this particular stage the core composition is 64.4\% He, 27.6\% C, 0.6\% N, and 7.4\% O by mass, but by the end of the helium burning
phase the remainder consists mostly of oxygen (83.6\% by mass) and carbon (12.7\%) without a trace of nitrogen. Prior to helium ignition
and on the RGB the carbon-to-nitrogen number ratio in the core is only $\approx 1/100$. At higher masses and up to 9 \msun\ nitrogen
is destroyed even more rapidly during the core helium-burning phase, and on the RGB the carbon-to-nitrogen number ratio in the core 
never rises above $\approx 1/40$. Should burning be terminated while on the RGB, nitrogen would dominate carbon in a 
helium-rich core.

We found that the DQ NLTT~16249A is part of a close double degenerate system.
Binary evolutionary scenarios \citep[e.g.,][]{nel2001} show that, normally, the mass of helium-core white dwarfs 
correlate with the double-degenerate orbital period and almost never exceed 0.4 \msun. Our spectral
decomposition suggests a DQ mass larger than $\approx 0.6$ \msun. Applying the Sch\"onberg-Chandrasekhar mass limit
to the case of a helium core and helium/hydrogen envelope, the core mass on the RBG could reach
$M_{c}/M=0.08$ prior to collapse and ignition, i.e., $M_{c}\approx 0.6$ \msun\ in the case of a $M=8$ \msun\ star.
In order to conceal a large helium-rich core the DQ white dwarf in NLTT~16249 would have to be the product
of such exceptional circumstances. 

Based on our spectral decomposition we find that
the stars have comparable luminosities with similar masses and cooling ages. In the lower mass range the cooling ages are
$t_{\rm\,DQ}=1.3$ Gyr and $t_{\rm\,DA}=1.5$ Gyr so that the DA may have preceded the DQ by $\la 200$ Myr, but in the higher
mass range the situation is reversed with $t_{\rm\,DQ}=2.6$ Gyr and $t_{\rm\,DA}=2.4$ Gyr.
In either case, the progenitors would follow each other off the main-sequence within
200 Myr or less. The maximum age differential could apply to initial masses of 3 and 4 \msun, or
in a lower mass range to initial masses of 2.0 and 2.2 \msun\ \citep[see][]{sch1992}.
Narrowing the age differential would accommodate even larger initial masses and we cannot exclude the possibility of a $M=8$ \msun\
progenitor for the DQ white dwarf although it is less likely than a 2-4 \msun\ progenitor. Moreover, a high-mass progenitor for the 
0.6-0.9 \msun\ DA white dwarf is also unlikely \citep{wei2000}. Therefore, NLTT~16249 is most probably a CO$+$CO system with the
DA forming first, and the DQ forming next during a common-envelope phase; the progenitor masses would be in the 2-4 \msun\ range.

The question of the source of nitrogen remains. So far, we considered the composition at the center of the star as representative of
the core. As pointed out by \citet{mac1998} in the case of oxygen, the material actually dredged-up is not located at the core center but,
instead, below the former helium-burning shell. Therefore, we propose that helium burning terminated prematurely during a common-envelope
phase leaving an abundance ratio C/N$\approx50$ off-center and below the former helium-burning shell in NLTT~16249A. 
In conclusion, the presence of nitrogen in the atmosphere of DQ white dwarfs would help distinguish normal DQ white dwarfs
that evolved in isolation and DQ white dwarfs that are the product of binary evolution.

\section{Summary}

We report the discovery of a peculiar DQ white dwarf in a close degenerate system ($P\la13$ hrs). 
The presence of
nitrogen in the atmosphere of the DQ also suggests the presence of unprocessed material in the core of the star, most
probably off-center and below the extinct helium-burning shell. Helium burning may have been interrupted
following a common-envelope event leaving a nitrogen-rich layer above a normal carbon/oxygen core.
Our spectral decomposition suggests a system mass in the 1.3 to 1.8 \msun\ range with cooling ages
between 1.3 and 2.6 Gyr depending on the mass.
The binary will merge within 9 Gyr for a system mass of 1.3 \msun, or 13 Gyr for 1.8 \msun\ following
\citet{rit1986}. New radial velocity measurements will be used to set a precise binary mass ratio and confirm
the evolutionary prospects of the binary NLTT~16249.

\acknowledgments

S.V. and A.K. are supported by GA AV grant numbers IAA300030908 and IAA301630901, respectively, and by GA \v{C}R grant number P209/10/0967.
This research has made use of the VizieR catalogue access tool (CDS, Strasbourg, France), and
of data products from the Two Micron All Sky Survey which is a joint project 
of the University of Massachusetts and the Infrared Processing and Analysis Center/California Institute 
of Technology, funded by the National Aeronautics and Space Administration and the National Science Foundation.

{\it Facilities:} \facility{KPNO}.


\begin{thebibliography}{}
\bibitem[Bauschlicher et al.(1988)]{bau1988} Bauschlicher, C.~W., Jr., Langhoff, S.~R., \& Taylor, P.~R.\ 1988, \apj, 332, 531 
\bibitem[Benvenuto \& Althaus(1999)]{ben1999} Benvenuto, O.~G., \& Althaus, L.~G.\ 1999, \mnras, 303, 30
\bibitem[Bessell(1990)]{bes1990} Bessell, M.~S.\ 1990, \pasp, 102, 1181 
\bibitem[Dufour et al.(2005)]{duf2005} Dufour, P., Bergeron, P., \& Fontaine, G.\ 2005, \apj, 627, 404 
\bibitem[Eggen(1968)]{egg1968} Eggen, O.~J. 1968, ApJS, 16, 97
\bibitem[Fontaine et al.(1984)]{fon1984} Fontaine, G., Villeneuve, B., Wesemael, F., \& Wegner, G.\ 1984, \apjl, 277, L61 
\bibitem[G{\"a}nsicke et al.(2010)]{gae2010} G{\"a}nsicke, B.~T., Koester, D., Girven, J., Marsh, T.~R., \& Steeghs, D.\ 2010, Science, 327, 188 
\bibitem[Garcia-Berro et al.(1997)]{gar1997} Garcia-Berro, E., Ritossa, C., \& Iben, I., Jr.\ 1997, \apj, 485, 765 
\bibitem[Golden(1967)]{gol1967} Golden, S.\ 1967, \jqsrt, 7, 225
\bibitem[Herwig et al.(1999)]{her1999} Herwig, F., Bl{\"o}cker, T., Langer, N., \& Driebe, T.\ 1999, \aap, 349, L5 
\bibitem[Karakas et al.(2009)]{kar2009} Karakas, A.~I., van Raai, M.~A., Lugaro, M., Sterling, N.~C., \& Dinerstein, H.~L.\ 2009, \apj, 690, 1130 
\bibitem[Kawka \& Vennes(2006)]{kaw2006} Kawka, A., \& Vennes, S.\ 2006, \apj, 643, 402 
\bibitem[Kawka \& Vennes(2011)]{kaw2011} Kawka, A., \& Vennes, S.\ 2011, \aap, 532, A7 
\bibitem[Koester(2009)]{koe2009} Koester, D.\ 2009, \aap, 498, 517 
\bibitem[Koester \& Knist(2006)]{koe2006} Koester, D., \& Knist, S.\ 2006, \aap, 454, 951 
\bibitem[Koester et al.(1982)]{koe1982} Koester, D., Weidemann, V., \& Zeidler, E.-M.\ 1982, \aap, 116, 147 
\bibitem[Lattanzio et al.(1996)]{lat1996} Lattanzio, J., Frost, C., Cannon, R., \& Wood, P.~R.\ 1996, \memsai, 67, 729 
\bibitem[MacDonald et al.(1998)]{mac1998} MacDonald, J., Hernanz, M., \& Jose, J.\ 1998, \mnras, 296, 523 
\bibitem[Nelemans et al.(2001)]{nel2001} Nelemans, G., Yungelson, L.~R., Portegies Zwart, S.~F., \& Verbunt, F.\ 2001, \aap, 365, 491 
\bibitem[Pelletier et al.(1986)]{pel1986} Pelletier, C., Fontaine, G., Wesemael, F., Michaud, G., \& Wegner, G.\ 1986, \apj, 307, 242 
\bibitem[Ram et al.(2006)]{ram2006} Ram, R.~S., Davis, S.~P., Wallace, L., et al.\ 2006, Journal of Molecular Spectroscopy, 237, 225 
\bibitem[Reddy et al.(2003)]{red2003} Reddy, R.~R., Nazeer Ahammed, Y., Rama Gopal, K., \& Baba Basha, D.\ 2003, \apss, 286, 419 
\bibitem[Ritter(1986)]{rit1986} Ritter, H.\ 1986, \aap, 169, 139 
\bibitem[Schaller et al.(1992)]{sch1992} Schaller, G., Schaerer, D., Meynet, G., \& Maeder, A.\ 1992, \aaps, 96, 269 
\bibitem[Salim \& Gould(2003)]{sal2003} Salim, S., \& Gould, A. 2003, ApJ, 582, 1011
\bibitem[Sauval \& Tatum(1984)]{sau1984} Sauval, A.~J., \& Tatum, J.~B.\ 1984, \apjs, 56, 193 
\bibitem[Sion et al.(1983)]{sio1983} Sion, E.~M., Greenstein, J.~L., Landstreet, J.~D., et al.\ 1983, \apj, 269, 253 
\bibitem[Skrutskie et al.(2006)]{skr2006} Skrutskie, M.~F., et al.\ 2006, \aj, 131, 1163
\bibitem[Tanabashi et al.(2007)]{tan2007} Tanabashi, A., Hirao, T., Amano, T., \& Bernath, P.~F.\ 2007, \apjs, 169, 472 
\bibitem[Vauclair \& Fontaine(1979)]{vau1979} Vauclair, G., \& Fontaine, G.\ 1979, \apj, 230, 563 
\bibitem[Vernet et al.(2011)]{ver2011} Vernet, J., Dekker, H., D'Odorico, S., et al.\ 2011, \aap, in press (2011; DOI: 10.1051/0004-6361/201117752)
\bibitem[Wallace(1962)]{wal1962} Wallace, L.\ 1962, \apjs, 7, 165 
\bibitem[Wegner(1984)]{weg1984a} Wegner, G.\ 1984, \apss, 104, 347 
\bibitem[Wegner \& Yackovich(1984)]{weg1984} Wegner, G., \& Yackovich, F.~H.\ 1984, \apj, 284, 257 
\bibitem[Weidemann(2000)]{wei2000} Weidemann, V.\ 2000, \aap, 363, 647 
\bibitem[Zeidler-K.T.(1987)]{zei1987} Zeidler-K.T., E.-M.\ 1987, \aaps, 68, 469 
\bibitem[Zeidler-K.T.~\& Koester(1982)]{zei1982} Zeidler-K.T., E.~M., \& Koester, D.\ 1982, \aap, 113, 173 

\end{thebibliography}
\end{document}